# The Equation of State of Nuclear Matter in Heavy Ion Collisions at CERN-SPS Energies from the Viewpoint of Relativistic Hydrodynamics

B.R. Schlei[1,2], D. Strottman[2] and N. Xu[3]

[1] Physics Division, P-25, Los Alamos National Laboratory, Los Alamos, USA
[2] Theoretical Division, DDT-DO, Los Alamos National Laboratory, Los Alamos, USA
[3] Nuclear Science Division, Lawrence Berkeley National Laboratory, Berkeley, USA

We present refits of single inclusive cross section data of mesons and baryons for central 158 $AGeV$ Pb+Pb collisions measured by the NA44 and NA49 Collaborations. In the theoretical approach we use the relativistic hydrodynamical code HYLANDER-C. We investigate several equations of state in their capability to describe the measured single inclusive momentum distributions. Based on the obtained fits we present results of the calculated Bose-Einstein correlation functions of identical pion pairs and discuss their features in comparison to measurements.

Contributed Poster to the "13[th] International Conference on
Ultra-Relativistic Nucleus-Nucleus Collisions: **Quark Matter 97**",
Tsukuba, Japan, December 01-05, 1997.

# The Model: HYLANDER-C

Improved version of **HYLANDER**.

U. Ornik, F. Pottag, R.M. Weiner,
Phys. Rev. Lett. <u>63</u>, 2641 (1989).

## Numerical Solution of the Hydrodynamical EULER Equations

1. $\partial_t E = -\vec{\nabla} \cdot \vec{M}$
2. $\partial_t M_i = -(\vec{\nabla} \cdot (M_i \vec{v}) + \partial_i P)$
3. $\partial_t B = -\vec{\nabla} \cdot (B \vec{v})$

## Direct Use of an Equation of State (EOS)

4. $P(\epsilon, n) = c^2(\epsilon, n) \, \epsilon$

## Useful Relations

(a) $P/\epsilon = c^2(\epsilon) = \dfrac{1}{\epsilon} \displaystyle\int_0^\epsilon c_o^2(\epsilon') \, d\epsilon'$

(b) $T(\epsilon) = T_c \exp\left[\displaystyle\int_{\epsilon_c}^{\epsilon} \dfrac{c_o^2(\epsilon') \, d\epsilon'}{(1 + c^2(\epsilon')) \, \epsilon'}\right]$

# Input for HYLANDER-C

## Type of EOS

Any type of EOS can be used.

1. in tabular form.
2. through parametrization.

   (a) K. Redlich, H. Satz,
       Phys. Rev. D33, 3747 (1986).   → I

   (b) C.M. Hung, E.V. Shuryak,
       Phys. Rev. Lett. 75, 4003 (1995).   → II

   (c) H. Sorge,
       "Soft transverse expansion in Pb(158 AGeV) on Pb collisions: preequilibrium motion or 1st order phase transition?",
       (1997) SUNY-NTG 97-1, nucl-th/9701012.   → III

## Initial Conditions

Any type of initial conditions can be used.

1. in tabular form.
2. through parametrization.

   (a) J. Bolz, U. Ornik, R.M. Weiner,
       Phys. Rev. C46, 2047 (1992).

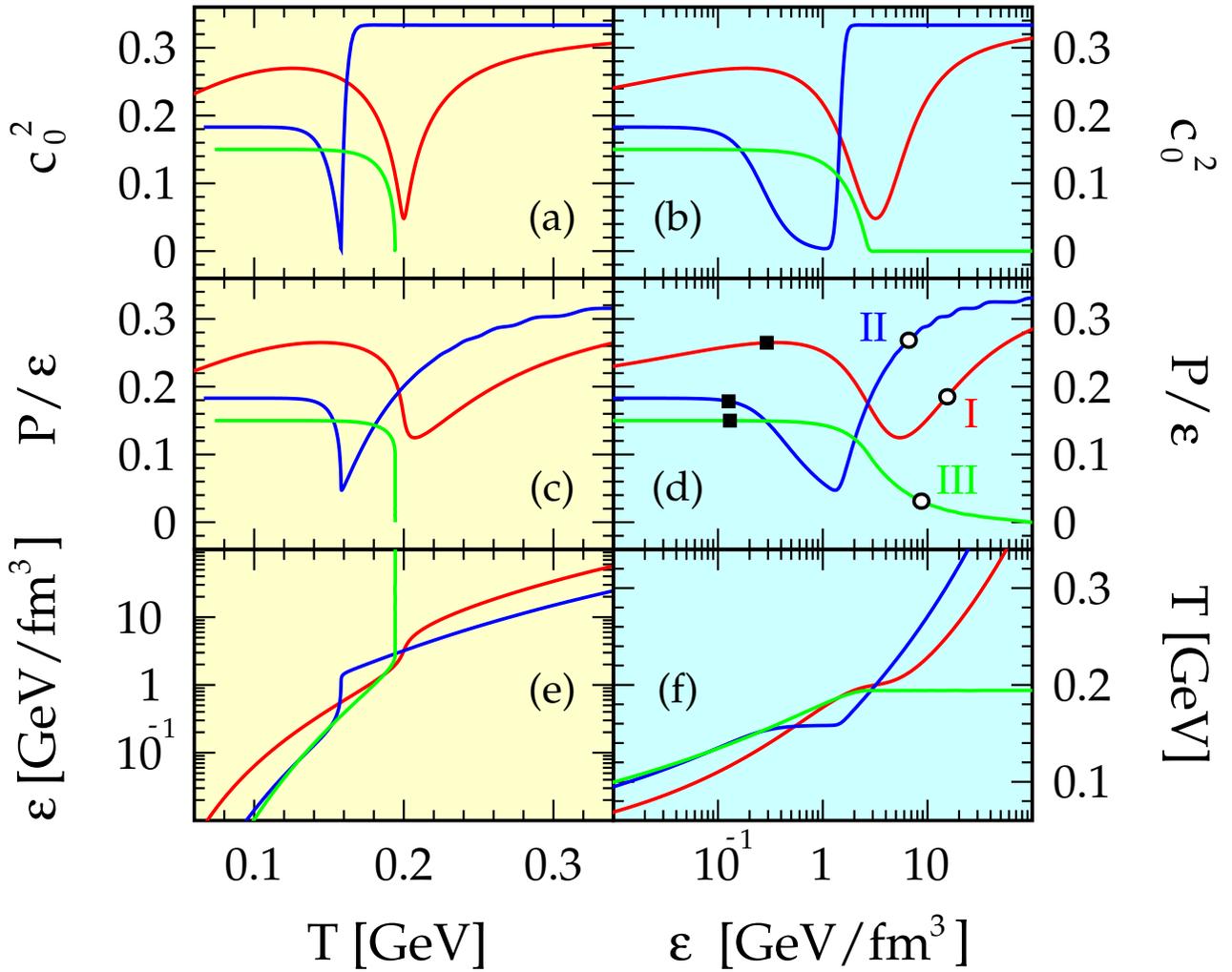

Energy density, $\epsilon$, ratio of pressure and energy density, $P/\epsilon$, speed of sound, $c_0^2$, and temperature, $T$, as functions of $T$ and/or $\epsilon$, for the equations of state EOS-I (red lines), EOS-II (blue lines), and EOS-III (green lines), respectively. The open circles in plot (d) correspond for each EOS to the starting values of $P/\epsilon$ with respect to the achieved initial maximum energy density $\epsilon_\Delta$ at transverse position $r_\perp = 0$, whereas the filled squares represent the final values of $P/\epsilon$ at breakup energy densities, $\epsilon_f$.

# Single Inclusive Distributions

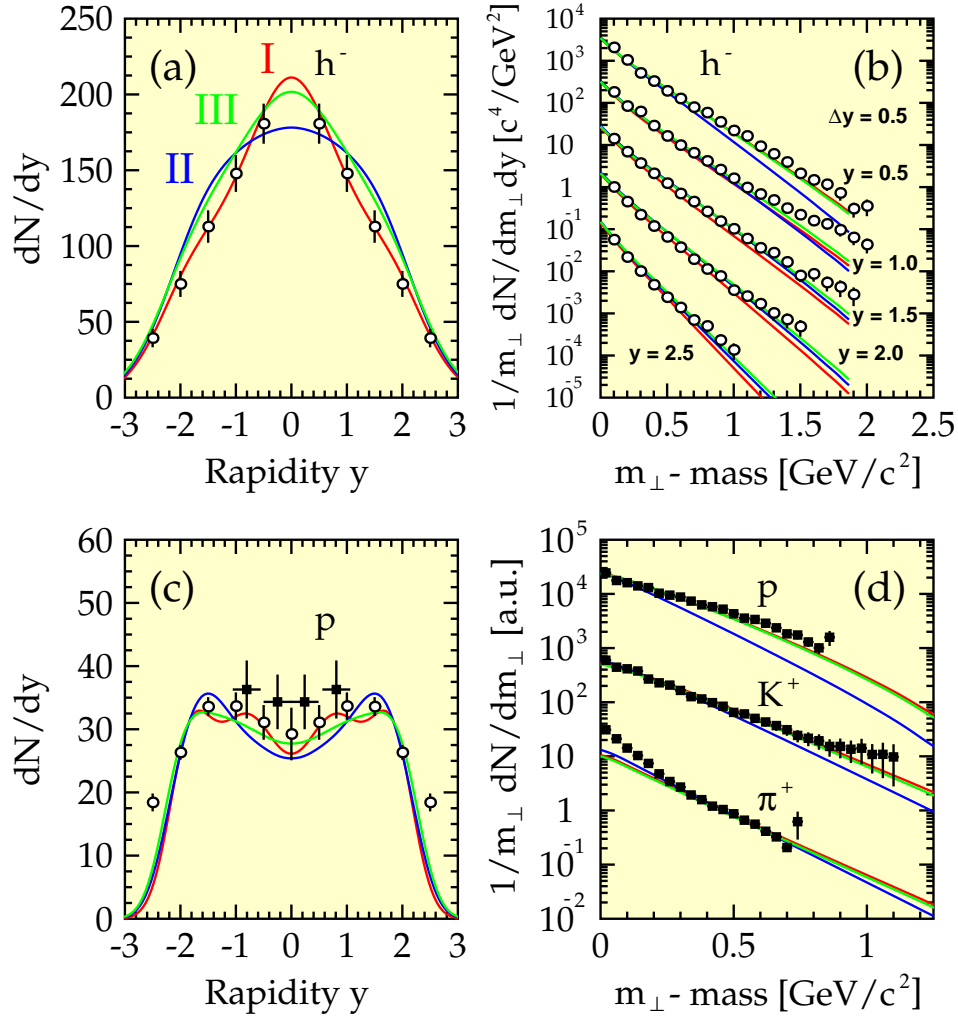

(a) Rapidity spectra and (b) transverse mass spectra, $1/m_\perp dN/dm_\perp dy$, of negative hadrons, $h^-$, (c) rapidity spectra of protons (without contributions from $\Lambda^0$ decay) and (d) transverse mass spectra, $1/m_\perp dN/dm_\perp$, of protons (including contributions from $\Lambda^0$ decay), $p$, positive kaons, $K^+$, and positive pions, $\pi^+$, respectively. The red (blue (green)) lines indicate the results of the calculations when using equation of state EOS-I (EOS-II (EOS-III)). The open circles represent preliminary data taken by the NA49 Collaboration, whereas the filled squares represent final data taken by the NA44 Collaboration.

# Bose-Einstein Correlations

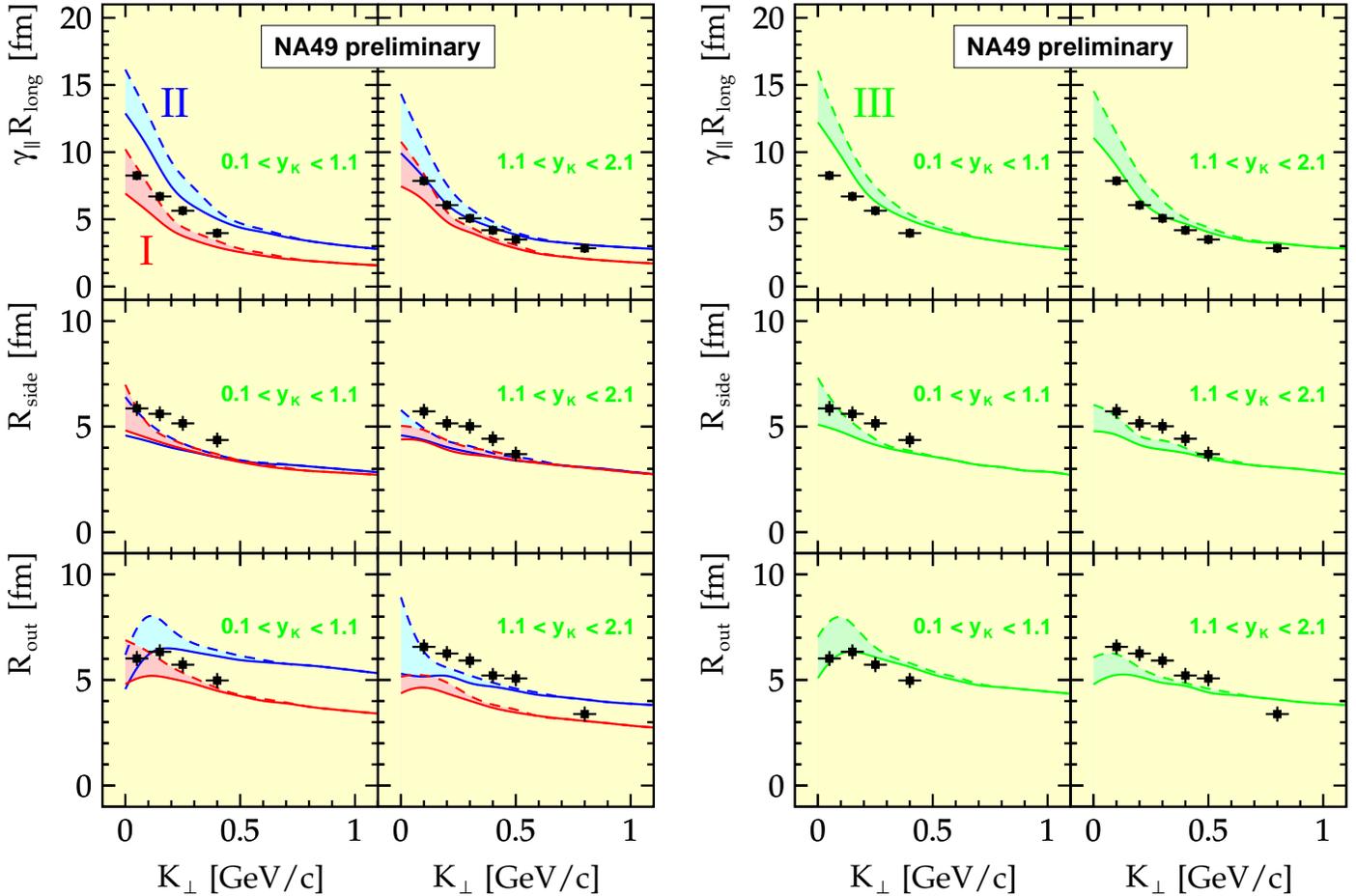

Inverse widths of BEC functions of identical negative pion pairs (including decay contributions from resonances) as functions of the average transverse momentum of the boson pair, $K_\perp$, in the indicated ranges of the particle pair rapidities, $y_K$, compared to preliminary data of the NA49 Collaboration. The solid lines indicate the inverse widths of BEC functions extracted from a Gaussian fit for the calculation using EOS-I (EOS-II (EOS-III)). The dashed lines are the true inverse widths of the correlation functions at their 68% level, and the colored zones reflect the theoretical uncertainties, when extracting the inverse widths.

# Results

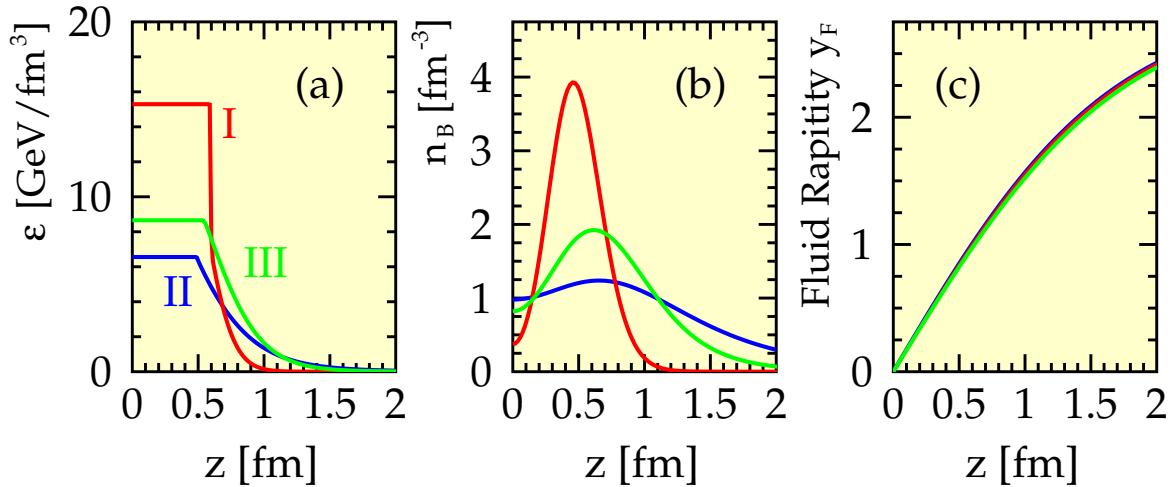

Initial distributions of (a) the energy density, $\epsilon$, (b) the baryon density, $n_B$, and (c) the fluid rapidity, $y_F$, plotted against the longitudinal coordinate $z$ at transverse position $r_\perp = 0$. The red (blue (green)) lines indicate the initial distributions for the calculation using EOS-I (EOS-II (EOS-III)).

## Special Features

1. FULL THREE-DIMENSIONAL SOLUTIONS
2. TRANSVERSE EXPANSION
3. EFFECTS OF RESONANCE DECAY
4. FREEZE-OUT TEMPERATURE, $T_f = 139 MeV$

## References

B.R. Schlei, Heavy Ion Phys. 5 (1997) 403.

B.R. Schlei, D. Strottman, and N. Xu,
    Los Alamos Preprint LA-UR-97-4230.